\documentclass[pr,onecolumn,10pt]{revtex4-2}
\usepackage[colorlinks,citecolor=blue,linkcolor=red]{hyperref}
\usepackage{amsmath,amssymb,graphicx,subfigure}
\usepackage{times}
\usepackage{subfigure}

\begin{document}

\title{Domain walls in fractional media}
\author{Shatrughna Kumar$^{1}$, Pengfei Li$^{2,3}$, and Boris A. Malomed$%
^{1,4}$}
\affiliation{$^{1}$Department of Physical Electronics, School of Electrical Engineering,
Faculty of Engineering, and Center for Light-Matter Interaction, Tel Aviv
University, P.O.B. 39040, Tel Aviv, Israel\\
$^{2}$Department of Physics, Taiyuan Normal University, Jinzhong, 030619, China\\
$^{3}$2Institute of Computational and Applied Physics, Taiyuan Normal University,
Jinzhong, 030619, China\\
$^4$Instituto de Alta Investigaci\'{o}n, Universidad de Tarapac\'{a},
Casilla 7D, Arica, Chile}

\begin{abstract}
Currently, much interest is drawn to the analysis of optical and matter-wave
modes supported by the fractional diffraction in nonlinear media. We predict
a new type of such states, in the form of domain walls (DWs) in the
two-component system of immiscible fields. Numerical study of the underlying
system of fractional nonlinear Schr\"{o}dinger equations demonstrates the
existence and stability of DWs at all values of the respective L\'{e}vy
index ($\alpha <2$) which determines the fractional diffraction, and at all
values of the XPM/SPM ratio $\beta $ in the two-component system above the
immiscibility threshold. The same conclusion is obtained for DWs in the
system which includes the linear coupling, alongside the XPM interaction
between the immiscible components. Analytical results are obtained for the
scaling of the DW's width. The DW solutions are essentially simplified in
the special case of $\beta =3$, as well as close to the immiscibility
threshold. In addition to symmetric DWs, asymmetric ones are constructed
too, in the system with unequal diffraction coefficients and/or different L%
\'{e}vy indices of the two components.
\end{abstract}

\maketitle


\section{Introduction}

The Schr\"{o}dinger equation with fractional spatial dispersion was
originally derived for the wave function of particles moving by L\'{e}vy
flights, using the Feynman-integral formulation of fundamental quantum
mechanics \cite{Lask1,Lask2}. While experimental realization of fractional
quantum mechanics has not been reported yet, it was proposed to emulate it
in terms of classical photonics, using the commonly known similarity of the
Schr\"{o}dinger equations and equations for the paraxial diffraction of
optical beams \cite{EXP3,PROP}. A universal method for the emulation of the
fractional diffraction is to use the basic 4\textit{f} configuration, which
makes it possible to perform the spatial Fourier transform of the beam,
apply the phase shift, which is tantamount to the action of the fractional
diffraction, by means of an appropriate phase plate, and finally transform
the beam back from the Fourier space \cite{EXP3}. In addition to that,
implementations of the fractional Schr\"{o}dinger equations were proposed in
Levy crystals \cite{EXP1} and polariton condensates \cite{EXP2}.

Theoretical studies initiated by the above-mentioned scheme were developed
in various directions, including the interplay of the fractional diffraction
with parity-time ($\mathcal{PT}$) symmetric potentials \cite{PTS}-\cite%
{ghosts}, propagation of Airy waves in the fractional geometry \cite%
{Yingji1,Yingji2}, and adding the natural Kerr nonlinearity to the
underlying setting, thus introducing fractional nonlinear Schr\"{o}dinger
equations (FNLSEs). The work with nonlinear models has produced many
predictions, such as the modulational instability of continuous waves (CWs)
\cite{Conti} and diverse types of optical solitons \cite{Jorge1}-\cite%
{review}. These are quasi-linear \textquotedblleft accessible solitons" \cite%
{Frac1,Frac2}, gap solitons maintained by lattice potentials \cite{Frac5a}-%
\cite{Frac5}, self-trapped vortices \cite{Frac6,Frac7}, multi-peak \cite%
{Frac8}-\cite{Frac11} and cluster \cite{Frac12} modes, fractional solitons
in discrete systems \cite{Frac13}, localized states featuring spontaneously
broken symmetry \cite{Frac15,Frac16,Frac17}, solitons in dual-core couplers
\cite{Frac18,Frac19}, solitary states supported by the quadratic
nonlinearity \cite{Thirouin,quadratic}, and dark modes \cite{we}. Also
studied were dissipative solitons in the fractional version of the complex
Ginzburg-Landau equation \cite{Frac14}. Many of these results were reviewed
in Ref. \cite{review}.

The objective of the present work is to introduce one-dimensional settings
for binary immiscible fields under the action of the fractional diffraction.
The immiscibility naturally gives rise to stable patterns in the form of
domain walls (DW), alias grain boundaries, which separate half-infinite
domains filled by the immiscible field components. In areas of traditional
physical phenomenology, DWs are well known as basic patterns in thermal
convection \cite{Manneville}-\cite{PLA}. Grain boundaries of a different
physical origin occur in various condensed-matter settings \cite{grain1}-%
\cite{grain3}. In optics, DWs were predicted and experimentally observed in
bimodal light propagation in fibers \cite{optical-DW,optical-DW2}. Similar
states were predicted in binary Bose-Einstein condensates (BECs), provided
that the inter-component repulsion is stronger than the self-repulsion of
each component, which provides for the immiscibility \cite{Mineev}-\cite%
{BEC-DW}.

The interplay of the two-component immiscibility, that maintains DWs, with
fractional diffraction may naturally appear in optics, considering the
fractional bimodal propagation of light in a self-defocusing spatial
waveguide. A similar model, based on a system of fractional Gross-Pitaevskii
equations (FGPEs) \cite{review}, may also naturally emerge in a binary BEC
composed of repulsively interaction particles which move by L\'{e}vy
flights. We construct DW solutions for coupled FNLSEs and verify their
stability by means of numerical methods. Some results -- in particular,
scaling relations which determine the DW's width as a function of basic
parameters of the system -- are obtained in an analytical form.

The paper is organized as follows. The model is formulated in Section 2,
which also includes analytical expressions for CW, i.e., spatially uniform
states, that may be linked by DW patterns, thus supporting them. Analytical
results for the DWs are collected in Section 3. Numerical results are
reported in Section 4, and the paper is concluded by Section 5.

\section{The model and CW states}

\subsection{Basic equations}

In terms of the optical bimodal propagation in the spatial domain, the
scaled system of coupled FNLSEs for amplitudes of copropagating
electromagnetic waves $u\left( x,z\right) $ and $v\left( x,z\right) $ with
orthogonal polarizations is
\begin{eqnarray}
i\frac{\partial u}{\partial z} &=&\frac{1}{2}\left( -\frac{\partial ^{2}}{%
\partial x^{2}}\right) ^{\alpha /2}u+(|u|^{2}+\beta |v|^{2})u-\lambda v,
\notag \\
i\frac{\partial v}{\partial z} &=&\frac{1}{2}\left( -\frac{\partial ^{2}}{%
\partial x^{2}}\right) ^{\alpha /2}v+(|v|^{2}+\beta |u|^{2})v-\lambda u,
\label{system}
\end{eqnarray}%
where $z$ is the propagation distance, $x$ is the transverse coordinate, and
the cubic terms, with normalized coefficients $1$ and $\beta >0$, represent,
respectively, the defocusing nonlinearity of the self-phase-modulation (SPM)
and cross-phase-modulation (XPM) types. The optical self-defocusing occurs,
in particular, in semiconductor waveguides \cite{semi}. In the BEC model,
the SPM and XPM terms represent repulsive interactions between two atomic
states in the binary condensate. In the latter case, the system of scaled
FGPEs is written in the form of Eq. (\ref{system}), with $z$ replaced by the
temporal variable, $t$.

In optics, two natural values of the XPM coefficient are $\beta =2$ for
components $u$ and $v$ representing circular polarizations of light, or $%
\beta =2/3$ in the case of linear polarizations \cite{Agrawal}. The value of
$\beta $ may be varied in broader limits (in particular, the case of $\beta
=3$ plays an essential role below) in photonic crystals \cite%
{PhotCryst,PhotCryst2}. In binary BEC, the effective XPM coefficient can be
readily adjusted by means of the Feshbach resonance \cite{Inguscio,Feshbach}.

In the case of orthogonal linear polarizations in optics [corresponding to $%
\beta =2/3$ in Eq. (\ref{system})], the nonlinear interaction between the
components includes, in addition to the XPM terms, also the four-wave mixing
(FWM), represented by terms $\left( 1/3\right) v^{2}u^{\ast }$ and $\left(
1/3\right) u^{2}v^{\ast }$ in FNLSEs (\ref{system}) for $u$ and $v$ (where $%
\ast $ stands for complex conjugate), although these terms are usually
suppressed by the phase-velocity-birefringence effect \cite{Agrawal}. In any
case, the FWM terms appearing in the optical system with orthogonal linear
polarizations are not relevant in the present context, as the condition of
the immiscibility of the two components holds only for $\beta >1$ [see Eq. (%
\ref{max}) below], eliminating the case of $\beta =2/3$. The optical system
with orthogonal circular polarizations corresponds, as said above, to $\beta
=2$, which admits the immiscibility, but the FWM terms do not appear in the
latter case. Normally, they do not appear either in the BEC model based on
the system of coupled FGPEs, therefore FWM terms are not considered here.

The linear-coupling terms with coefficient $\lambda \geq 0$ in Eq. (\ref%
{system}) account for mixing between the optical modes, or between the two
atomic states in BEC. In the former case, the linear mixing between circular
polarizations may be imposed by the birefringence \cite{Agrawal}, and in the
latter case the mutual conversion of atomic states in BEC\ may be driven by
resonant radiofrequency radiation \cite{radio}.

The fractional-diffraction operator with a positive L\'{e}vy index (LI) $%
\alpha $ is defined as the \textit{Riesz derivative }\cite%
{Agrawal-Riesz,Baleanu,Riesz},
\begin{gather}
\left( -\frac{\partial ^{2}}{\partial x^{2}}\right) ^{\alpha /2}u(x)\equiv
\notag \\
=\frac{1}{2\pi }\int_{-\infty }^{+\infty }|p|^{\alpha }dp\int_{-\infty
}^{+\infty }d\xi e^{ip(x-\xi )}u(\xi )\equiv \frac{1}{\pi }\int_{0}^{+\infty
}p^{\alpha }dp\int_{-\infty }^{+\infty }d\xi \cos \left( p(x-\xi \right)
)u(\xi ),  \label{Riesz}
\end{gather}%
which is built as the juxtaposition of the direct and inverse Fourier
transform, with the fractional diffraction acting at the intermediate stage.
While there are different definitions of fractional derivatives, this one
naturally appears in quantum mechanics \cite{Lask1,Lask2} and optics \cite%
{EXP3}. Normally, the LI takes values $1<\alpha \leq 2$, but, in the case of
the self-defocusing sign of the nonlinearity, when the system is not subject
to the wave collapse (implosion driven by self-attraction), it is also
possible to consider values $0<\alpha \leq 1$. The usual (non-fractional)
diffraction naturally corresponds to $\alpha =2$ in Eq. (\ref{system}).

Stationary solutions to Eqs. (\ref{system}) with propagation constant $k<0$
are looked for as%
\begin{equation}
\left\{ u\left( x,z\right) ,v\left( x,z\right) \right\} =e^{ikz}\left\{
U(x),V(x)\right\} ,  \label{uv}
\end{equation}%
where $U(x)$ and $V(x)$ are real functions which satisfy the following
system of equations:%
\begin{eqnarray}
kU+\frac{1}{2}\left( -\frac{\partial ^{2}}{\partial x^{2}}\right) ^{\alpha
/2}U+(U^{2}+\beta V^{2})U-\lambda V &=&0,  \notag \\
kV+\frac{1}{2}\left( -\frac{\partial ^{2}}{\partial x^{2}}\right) ^{\alpha
/2}V+(V^{2}+\beta U^{2})V-\lambda U &=&0.  \label{UV}
\end{eqnarray}%
The energy (Hamiltonian) of the stationary state (\ref{UV}) with the Riesz
derivatives defined as per Eq. (\ref{Riesz}) is%
\begin{gather}
E=\frac{1}{4\pi }\int_{-\infty }^{+\infty }dx\int_{-\infty }^{+\infty }d\xi
\int_{0}^{+\infty }p^{\alpha }dp\int_{-\infty }^{+\infty }d\xi \cos \left(
p(x-\xi \right) )\left[ U(x)U(\xi )+V(x)V(\xi )\right]  \notag \\
+\int_{-\infty }^{+\infty }dx\left[ \frac{1}{4}\left( U^{4}+V^{4}+2\beta
U^{2}V^{2}\right) -\lambda UV\right] .  \label{E}
\end{gather}

Stability of stationary solutions, obtained in the form of expression (\ref%
{uv}), against small perturbations was investigated by means of the usual
approach, looking for the perturbed solution as%
\begin{eqnarray}
u\left( x,z\right) &=&e^{ikz}\left[ U(x)+e^{\gamma z}a(x)+e^{\gamma ^{\ast
}z}b^{\ast }(x)\right] ,  \notag \\
v\left( x,z\right) &=&e^{ikz}\left[ V(x)+e^{\gamma z}c(x)+e^{\gamma ^{\ast
}z}d^{\ast }(x)\right] ,  \label{pert}
\end{eqnarray}%
where $\left\{ a(x),b(x),c(x),d(x)\right\} $ are components of an eigenmode
of infinitesimal perturbations, and $\gamma $ is the respective eigenvalue
(which may be a complex number). The substitution of the perturbed
expression (\ref{pert}) in Eq. (\ref{system}) and linearization leads to the
system of coupled equations,%
\begin{eqnarray}
\left( -k+i\gamma \right) a &=&\frac{1}{2}\left( -\frac{\partial ^{2}}{%
\partial x^{2}}\right) ^{\alpha /2}a+\left( 2U^{2}+\beta V^{2}\right)
a+U^{2}b+\beta UV(c+d)-\lambda c,  \notag \\
\left( -k-i\gamma \right) b &=&\frac{1}{2}\left( -\frac{\partial ^{2}}{%
\partial x^{2}}\right) ^{\alpha /2}b+\left( 2U^{2}+\beta V^{2}\right)
b+U^{2}a+\beta UV(c+d)-\lambda d,  \notag \\
&&  \label{linearized} \\
\left( -k+i\gamma \right) c &=&\frac{1}{2}\left( -\frac{\partial ^{2}}{%
\partial x^{2}}\right) ^{\alpha /2}c+\left( 2U^{2}+\beta V^{2}\right)
c+U^{2}d+\beta UV(a+b)-\lambda a,  \notag \\
\left( -k-i\gamma \right) d &=&\frac{1}{2}\left( -\frac{\partial ^{2}}{%
\partial x^{2}}\right) ^{\alpha /2}d+\left( 2U^{2}+\beta V^{2}\right)
d+U^{2}c+\beta UV(a+b)-\lambda b.  \notag
\end{eqnarray}%
The underlying DW solution is stable if numerical solution of Eq. (\ref%
{linearized}) yields solely imaginary eigenvalues, with zero real parts.

\subsection{Continuous-wave (CW) solutions and the immiscibility condition}

The spatially uniform version of Eq. (\ref{UV}), with $U,V=\mathrm{const}$,\
gives rise to two asymmetric (partly immiscible, with $U\neq V$) CW
solutions, labeled by subscripts $+$ and $-$, which are mirror images of
each other:%
\begin{eqnarray}
\left\{
\begin{array}{c}
U_{+} \\
V_{+}%
\end{array}%
\right\} &=&\frac{1}{\sqrt{2}}\left\{
\begin{array}{c}
\sqrt{-\frac{k}{2}+\frac{\lambda }{\beta -1}}+\sqrt{-\frac{k}{2}-\frac{%
\lambda }{\beta -1}} \\
\sqrt{-\frac{k}{2}+\frac{\lambda }{\beta -1}}-\sqrt{-\frac{k}{2}-\frac{%
\lambda }{\beta -1}}%
\end{array}%
\right\} ,  \label{CW+} \\
\left\{
\begin{array}{c}
U_{-} \\
V_{-}%
\end{array}%
\right\} &=&\frac{1}{\sqrt{2}}\left\{
\begin{array}{c}
\sqrt{-\frac{k}{2}+\frac{\lambda }{\beta -1}}-\sqrt{-\frac{k}{2}-\frac{%
\lambda }{\beta -1}} \\
\sqrt{-\frac{k}{2}+\frac{\lambda }{\beta -1}}+\sqrt{-\frac{k}{2}-\frac{%
\lambda }{\beta -1}}%
\end{array}%
\right\} .  \label{CW-}
\end{eqnarray}%
$\allowbreak $Note that the total density of solutions (\ref{CW+}) and (\ref%
{CW-}) is%
\begin{equation}
U_{+}^{2}+V_{+}^{2}=U_{-}^{2}+V_{-}^{2}=-k.  \label{density}
\end{equation}%
While in the limit of $\lambda =0$ (no linear mixing), the obvious CW states
are completely immiscible, with $V_{+}=U_{-}=0$, the partly immiscible
states given by Eqs. (\ref{CW+}) and (\ref{CW-}) were found only recently in
Ref. \cite{PLA}. Parallel to the asymmetric CW states (\ref{CW+}) and (\ref%
{CW-}) there is the mixed (symmetric) one, with%
\begin{equation}
U_{0}=V_{0}=\sqrt{(\lambda -k)/(1+\beta )}.  \label{CW0}
\end{equation}

For given $k$, i.e., for given CW density [see Eq. (\ref{density})], CW
states (\ref{CW+}) and (\ref{CW-}) exist under the following condition:%
\begin{equation}
\beta -1>\left( \beta -1\right) _{\mathrm{immisc}}\equiv 2\lambda /|k|.
\label{max}
\end{equation}%
In the absence of the linear mixing, $\lambda =0$, Eq. (\ref{max}) amounts
to the commonly known immiscibility condition \cite{Mineev}, $\beta >\beta _{%
\mathrm{immisc}}=1$. At $\lambda >0$, Eq. (\ref{max}) demonstrates that the
linear mixing pushes the immiscibility threshold to higher values, as was
first demonstrated in Ref. \cite{Merhasin} under normalization condition $%
k=-1$. Precisely at the threshold, i.e., at $\beta -1=2\lambda /|k|$, Eqs. (%
\ref{CW+}), (\ref{CW-}) and (\ref{CW0}) yield the following magnitude of the
CW fields,%
\begin{equation}
U_{\mathrm{thresh}}=V_{\mathrm{thresh}}=\sqrt{\lambda /(\beta -1)}.
\label{threshold}
\end{equation}

As shown in Ref. \cite{Merhasin}, the meaning of immiscibility condition (%
\ref{max}) can be understood in terms of energy: at $\beta -1>\left( \beta
-1\right) _{\mathrm{immisc}}$, for given density of the CW state [see Eq. (%
\ref{density})], the energy density of the partly immiscible state,
determined as per the second line of Eq. (\ref{E}), is lower than that of
the mixed one (\ref{CW0}), hence the asymmetric CW solutions (\ref{CW+}) and
(\ref{CW-}) play the role of the system's ground state, while the mixed CW
state (\ref{CW0}) is unstable. On the other hand, at $\beta -1<\left( \beta
-1\right) _{\mathrm{immisc}}$ the mixed CW state is the only existing one,
being the (stable) ground state in that case.

\section{Domain-wall (DW) solutions: Analytical findings}

DW states exist when Eq. (\ref{UV}) maintains the (partly) immiscible CW
states, as given by Eqs. (\ref{CW+}) and (\ref{CW-}). The DW solution links
two different CW states, that fill the space at $x\rightarrow \pm \infty $,
according to the following boundary conditions:
\begin{eqnarray}
\lim_{x\rightarrow +\infty }\left\{
\begin{array}{c}
U(x) \\
V(x)%
\end{array}%
\right\} &=&\left\{
\begin{array}{c}
U_{-} \\
V_{-}%
\end{array}%
\right\} ,  \notag \\
&&  \label{link} \\
\lim_{x\rightarrow -\infty }\left\{
\begin{array}{c}
U(x) \\
V(x)%
\end{array}%
\right\} &=&\left\{
\begin{array}{c}
U_{+} \\
V_{+}%
\end{array}%
\right\} ,  \notag
\end{eqnarray}

An essential fact is that, in the particular case of $\beta =3$, the system
of two equations (\ref{UV}) can be exactly reduced to a single equation, by
the substitution of%
\begin{equation}
\left\{
\begin{array}{c}
U(x) \\
V(x)%
\end{array}%
\right\} =\frac{1}{2}\left\{
\begin{array}{c}
\sqrt{\lambda -k}-W(x), \\
\sqrt{\lambda -k}+W(x),%
\end{array}%
\right\}  \label{ansatz}
\end{equation}%
where $W(x)$ is a real odd function of $x$ satisfying the equation%
\begin{equation}
(k+\lambda )W+\frac{1}{2}\left( -\frac{\partial ^{2}}{\partial x^{2}}\right)
^{\alpha /2}W+W^{3}=0,  \label{W}
\end{equation}%
which is supplemented by the boundary conditions%
\begin{equation}
\lim_{x\rightarrow \pm \infty }W(x)=\pm \frac{1}{2}\sqrt{-k-\lambda }
\label{lim}
\end{equation}%
[note that it follows from Eq. (\ref{max}) with $\beta =3$ that values $%
\sqrt{-k-\lambda }$ in Eq. (\ref{lim}) are real].

In the case of the fractional diffraction (for $\alpha <2$), Eqs. (\ref%
{ansatz}) and (\ref{W}) represent a new result for $\beta =3$, while in the
case of the usual diffraction, i.e., at $\alpha =2$, the respective solution
was recently found in a fully explicit form \cite{PLA}:
\begin{equation}
\left\{
\begin{array}{c}
U(x) \\
V(x)%
\end{array}%
\right\} _{\alpha =2,\beta =3}=\frac{1}{2}\left\{
\begin{array}{c}
\sqrt{-k+\lambda }-\sqrt{-k-\lambda }\tanh \left( \sqrt{-k-\lambda }x\right)
\\
\sqrt{-k+\lambda }+\sqrt{-k-\lambda }\tanh \left( \sqrt{-k-\lambda }x\right)%
\end{array}%
\right\} .  \label{exact}
\end{equation}%
The existence of the relevant solution to Eq. (\ref{W}) with $\alpha <2$ is
corroborated by the numerical results for dark solitons as solutions of the
FNLSE, which were reported (in a different context) in Ref. \cite{we}.
Furthermore, as $\left( k+\lambda \right) $ is the single control parameter
in Eq. (\ref{W}), an exact property of the solutions with all values of $%
\alpha $ is a scaling relation for the DW's width, $L$:%
\begin{equation}
L\sim \left( -k-\lambda \right) ^{-1/\alpha }.  \label{L}
\end{equation}%
In particular, Eq. (\ref{L}) agrees with the exact solution (\ref{exact}) in
the case of $\alpha =2$.

An approximate scaling relation for the DW can be constructed in the case
when propagation constant $k$ is taken close to the threshold value (\ref%
{max}), i.e., setting%
\begin{equation}
k=-\frac{2\lambda }{\beta -1}-q,  \label{q}
\end{equation}%
with
\begin{equation}
0<q\ll 2\lambda /(\beta -1).  \label{small-q}
\end{equation}%
In this case, Eqs. (\ref{link}) and (\ref{CW+}), (\ref{CW-}) show that the
DW links boundary values with a small difference between them,%
\begin{equation}
\left\{ U(x=\pm \infty ),V(x=\pm \infty )\right\} \approx \left\{ \sqrt{%
\lambda /(\beta -1)}\mp \sqrt{q}/2,\sqrt{\lambda /(\beta -1)}\pm \sqrt{q}%
/2\right\} ,  \label{a}
\end{equation}%
where the main term is the same as in Eq. (\ref{threshold}). Further,
straightforward analysis of Eq. (\ref{UV}) demonstrates that, under
condition (\ref{small-q}), the DW's width scales with the variation of $q$ as%
\begin{equation}
L\sim q^{-1/\alpha },  \label{WDS}
\end{equation}%
cf. Eq. (\ref{L}).

Finally, in the case of $\lambda =0$ in Eq. (\ref{UV}), the near-threshold
case is defined, instead of Eq. (\ref{small-q}), simply as

\begin{equation}
0<\beta -1\ll 1,  \label{<<}
\end{equation}%
see Eq. (\ref{max}). In this case, the analysis of Eq. (\ref{UV}) leads to
the following asymptotic scaling relation for the DW's width,%
\begin{equation}
L\sim \left( \beta -1\right) ^{-1/\alpha },  \label{beta-1}
\end{equation}%
cf. Eqs. (\ref{L}) and (\ref{beta-1}).

In the case of the normal diffraction, $\alpha =2$, the situation valid
under condition (\ref{<<}) was considered in Ref. \cite{optical-DW}, where
the scaling was obtained in the form of $L\sim (\beta -1)^{-1/2}$, cf. Eq. (%
\ref{beta-1}). Also in agreement with Eq. (\ref{beta-1}), DWs do not exist
in the system with the Manakov's nonlinearity \cite{Manakov}, $\beta =1$.

\section{Numerical results}

Numerical solutions of Eq. (\ref{UV}) for stationary DWs were produced by
means of the well-known Newton's conjugate-gradient method \cite{JYang}. The
results are presented below separately for the systems without and with the
linear coupling ($\lambda =0$ and $\lambda >0$, respectively), and also, in
a brief form for an asymmetric generalization of Eq. (\ref{UV}), with
unequal diffraction coefficients and/or values of the LI for components $U$
and $V$.

\subsection{The system without the linear coupling ($\protect\lambda =0$)}

\begin{figure}[tbp]
\begin{center}
\includegraphics[width=0.60\textwidth]{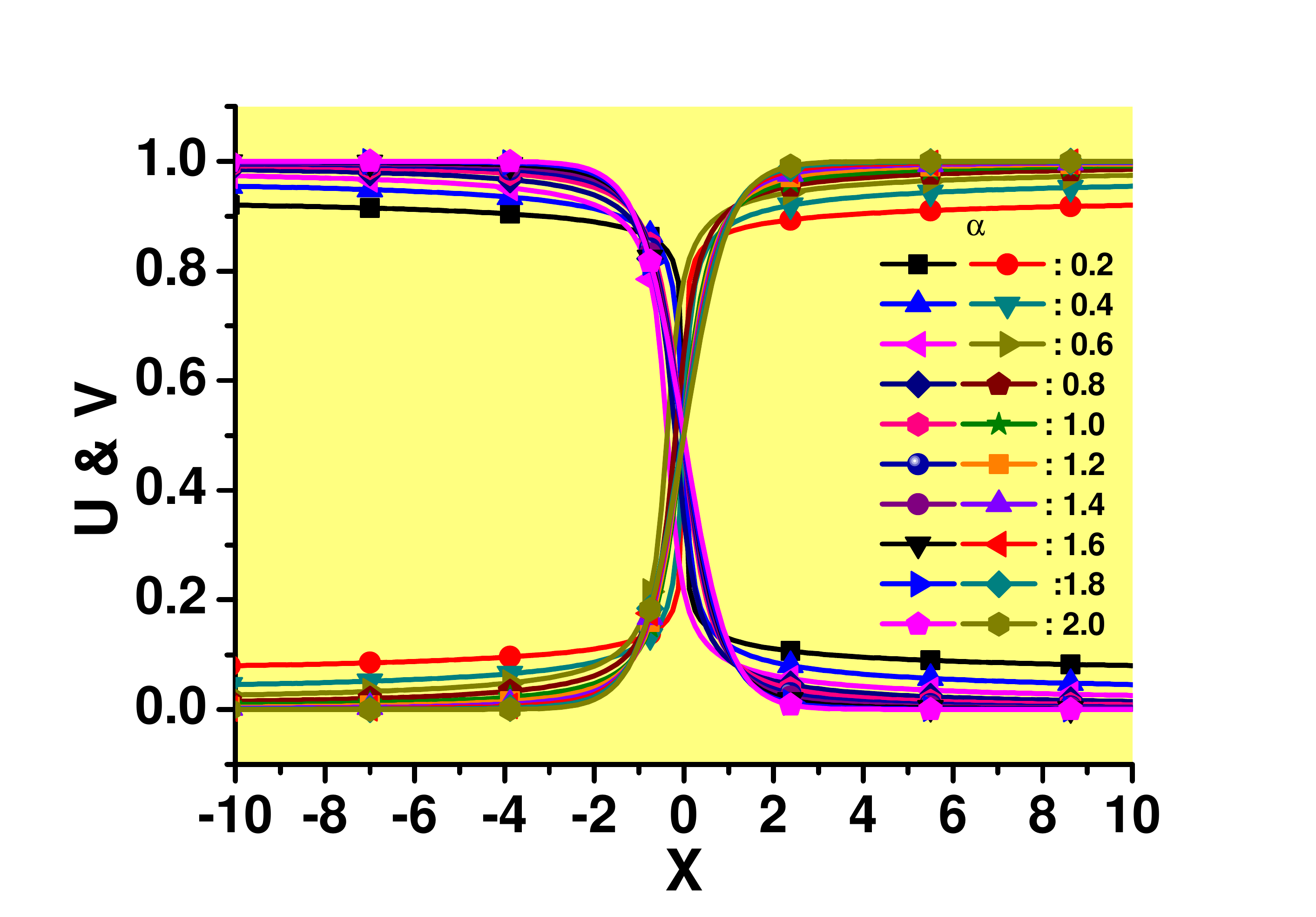}
\end{center}
\caption{A set of stationary profiles of the two components of the
numerically generated DW (domain-wall) solutions of Eq. (\protect\ref{UV})
with $\protect\lambda =0$, $\protect\beta =3$, $k=-1$ and indicated values
of the LI (L\'{e}vy index) varying from $\protect\alpha =2$ (which
corresponds to the normal non-fractional diffraction) up to $\protect\alpha %
=0.2$.}
\label{fig1}
\end{figure}

First, we consider the basic system of stationary equations (\ref{UV}) with $%
\lambda =0$. Results produced by the numerical solution of this system are
summarized in Fig. \ref{fig1} for $\beta =3$, because this value of the XPM
coefficient, as shown above, simplifies the system, allowing one to reduce
it to the single equation (\ref{W}). In this and similar figures, the
propagation constant is fixed as $k=-1$, which is always possible by means
of rescaling. It is seen that the DW patterns are truly robust ones, as the
variation of the LI in broad limits, from $\alpha =2$ up to $\alpha =0.2$,
produces relatively mild changes in the shape of the DWs, which persist, as
solutions of Eq. (\ref{UV}), at all values of the LI (numerical results are
not presented for very small values of $\alpha $, as the numerical method
encounters technical problems in that case).

\begin{figure}[tbp]
\begin{center}
\subfigure[]{\includegraphics[scale=0.9]{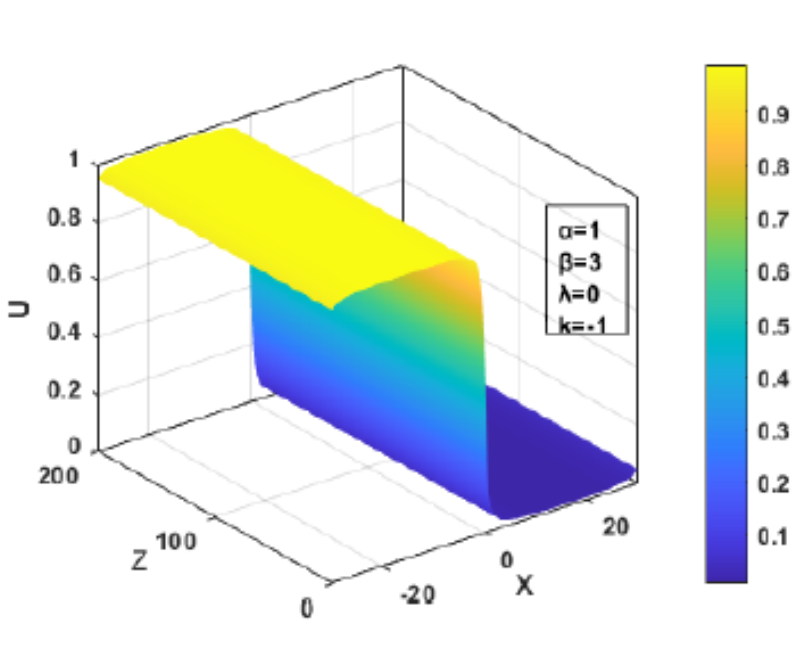}} \subfigure[]{%
\includegraphics[scale=0.9]{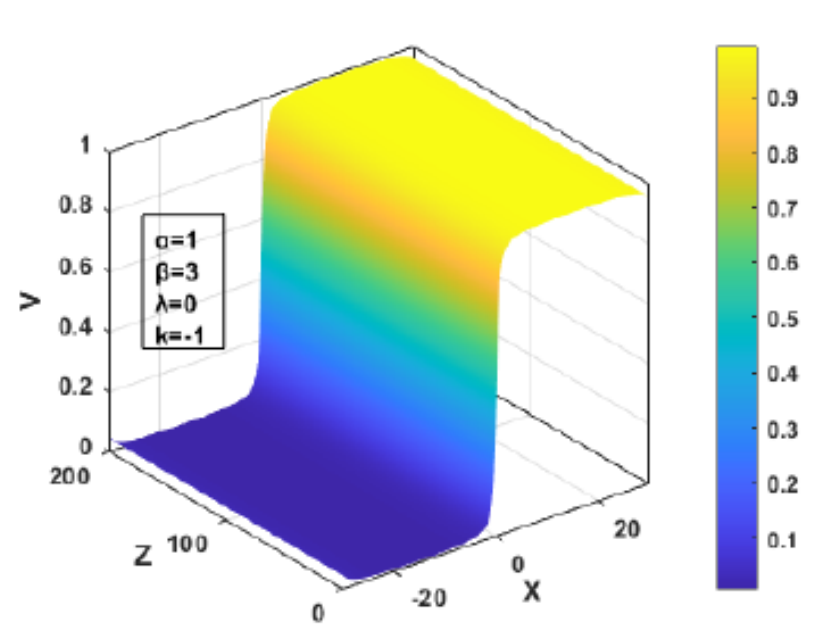}}
\end{center}
\caption{An example of the stable evolution of the two components of the DW
[shown in panels (a) and (b)] with $k=-1$, produced by simulations of Eq. (%
\protect\ref{system}) with $\protect\alpha =1$, $\protect\beta =3$, and $%
\protect\lambda =0$. The stability of the DW is corroborated by a purely
stable spectrum of eigenvalues produced, for the same DW state, by numerical
solution of Eq. (\protect\ref{linearized}) (not shown here).}
\label{fig2}
\end{figure}

Furthermore, the numerical solution of the system of linearized equations (%
\ref{linearized}) for small perturbations produces completely stable spectra
of eigenvalues for all stationary DW patterns (not shown here in detail, as
they do not exhibit essential peculiarities). The stability of the DWs was
also corroborated by direct simulations of the system of underlying
equations (\ref{system}), see a typical examples displayed in Fig. \ref{fig2}
for $\alpha =1$.

\begin{figure}[tbp]
\begin{center}
\includegraphics[width=0.60\textwidth]{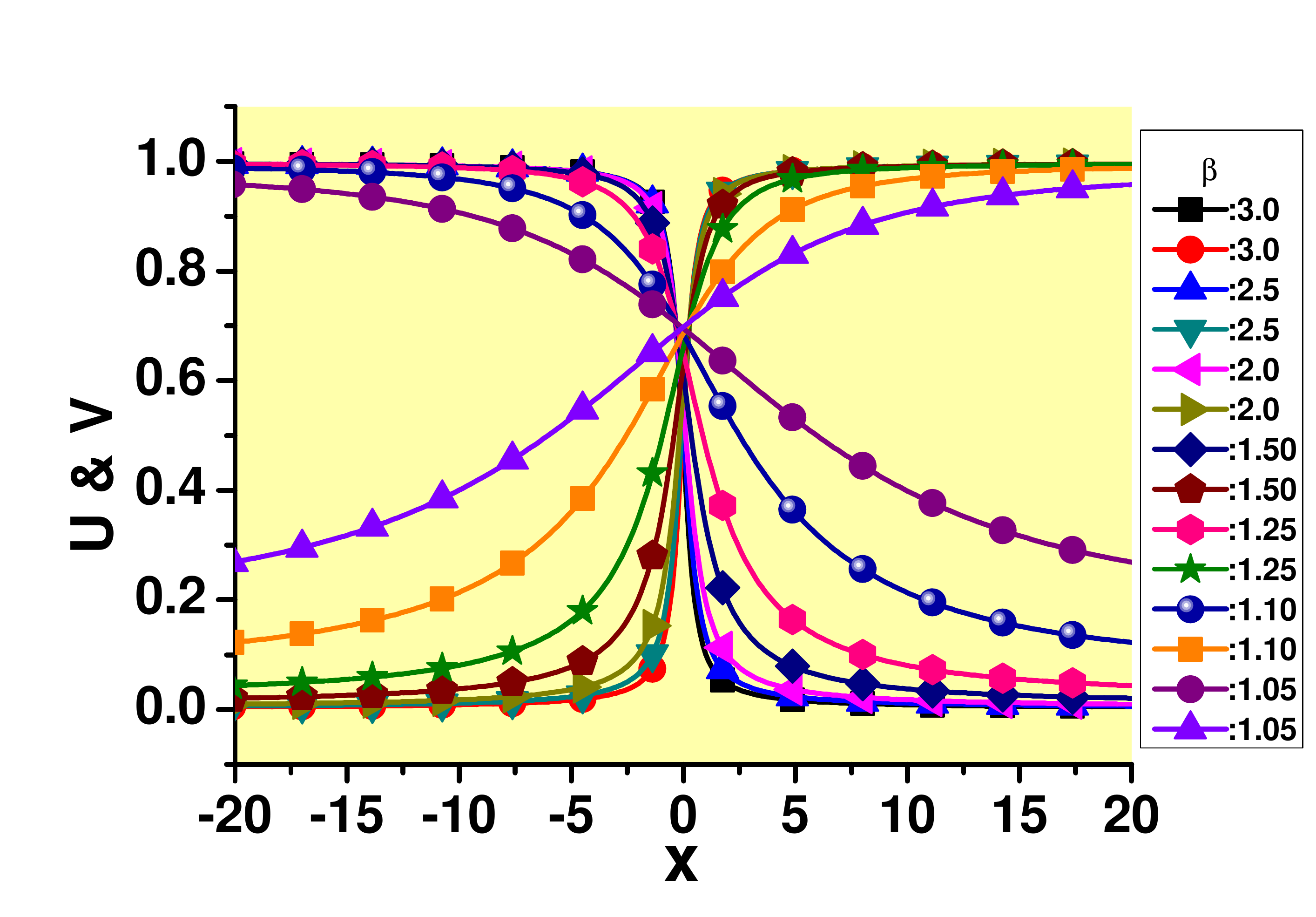}
\end{center}
\caption{A set of DW solutions with $k=-1$, produced by Eq. (\protect\ref{UV}%
) with $\protect\lambda =0$, $\protect\alpha =1$ and values of XPM
coefficient varying in interval $1.05\leq \protect\beta \leq 3$. All the
solutions are stable, according to the calculation of eigenvalues [see Eq. (%
\protect\ref{linearized})] and direct simulations of Eq. (\protect\ref%
{system}).}
\label{fig3}
\end{figure}

Similar results, i.e., a family of stable DW solutions, are produced by the
numerical analysis of Eqs. (\ref{system}), (\ref{UV}), and (\ref{linearized}%
), for values of the XPM coefficient $\beta \neq 3$, as shown by a set of
profiles in Fig. \ref{fig3} for $\alpha =1$ (it is a typical value of the LI
corresponding to the fractional diffraction).

\begin{figure}[tbp]
\begin{center}
\includegraphics[width=0.80\textwidth]{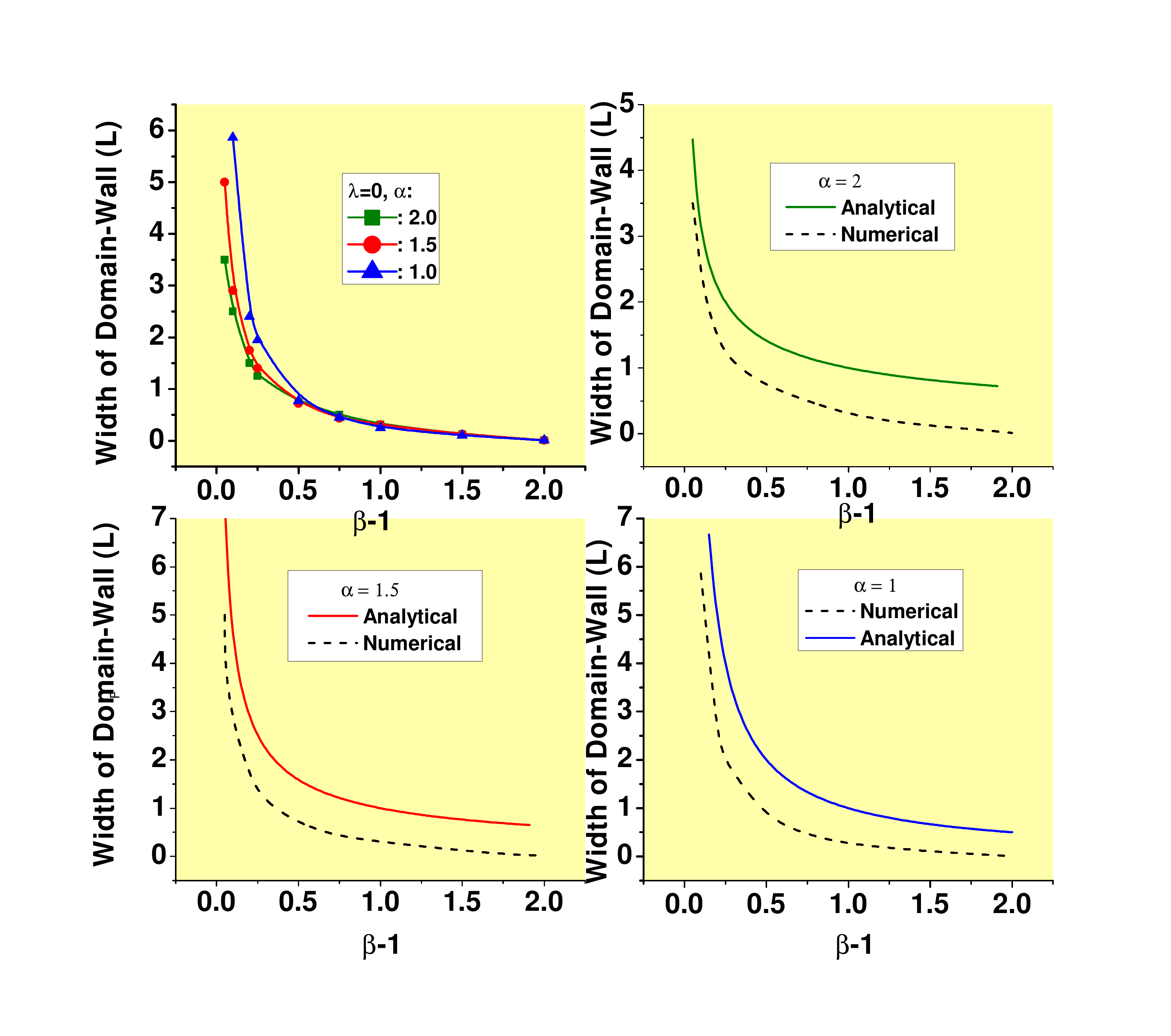}
\end{center}
\caption{The top left panel: Width $L$ of the DW for $k=-1$, $\protect%
\lambda =0$, and three different values of the LI, $\protect\alpha =2$
(which corresponds to the normal non-fractional diffraction), $\protect%
\alpha =1.5$, and $\protect\alpha =1,$ vs. the proximity, $\left( \protect%
\beta -1\right) $, to the DW's existence threshold, $\protect\beta =1$. The
other panels show the comparison of the respective numerically found curves,
$L(\protect\beta -1)$, to the asymptotic analytically predicted scaling
relation (\protect\ref{beta-1}).}
\label{fig4}
\end{figure}

The set of DW profiles is displayed in Fig. \ref{fig3} for $\beta \geq 1.05$%
, as, for very small values of $\beta -1$, the width of the DW diverges, in
accordance with the scaling relation given, for $\lambda =0$, by Eq. (\ref%
{beta-1}). For different values of $\alpha $ the DW families are
characterized by dependences of their width $L$ on $(\beta -1)$, as shown in
the top left panel of Fig. \ref{fig4}. The width was identified, from the
numerical solution, as the distance between points where $U(x)$ and $V(x)$
take values $1/2$, i.e., half of the asymptotic values $U(x\rightarrow
-\infty )=V(x=+\infty )=1$ for $k=-1$, see Eqs. (\ref{CW+}) and (\ref{CW-}).
In the same figure \ref{fig4}, the numerically found dependences are
compared to the analytically predicted asymptotic scaling relations given by
Eq. (\ref{beta-1}). It is seen that the prediction is indeed very close to
the numerical results for sufficiently small values of $\left( \beta
-1\right) $.

\subsection{The system including the linear coupling ($\protect\lambda >0$)}

\begin{figure}[tbp]
\subfigure[]{\includegraphics[scale=0.3]{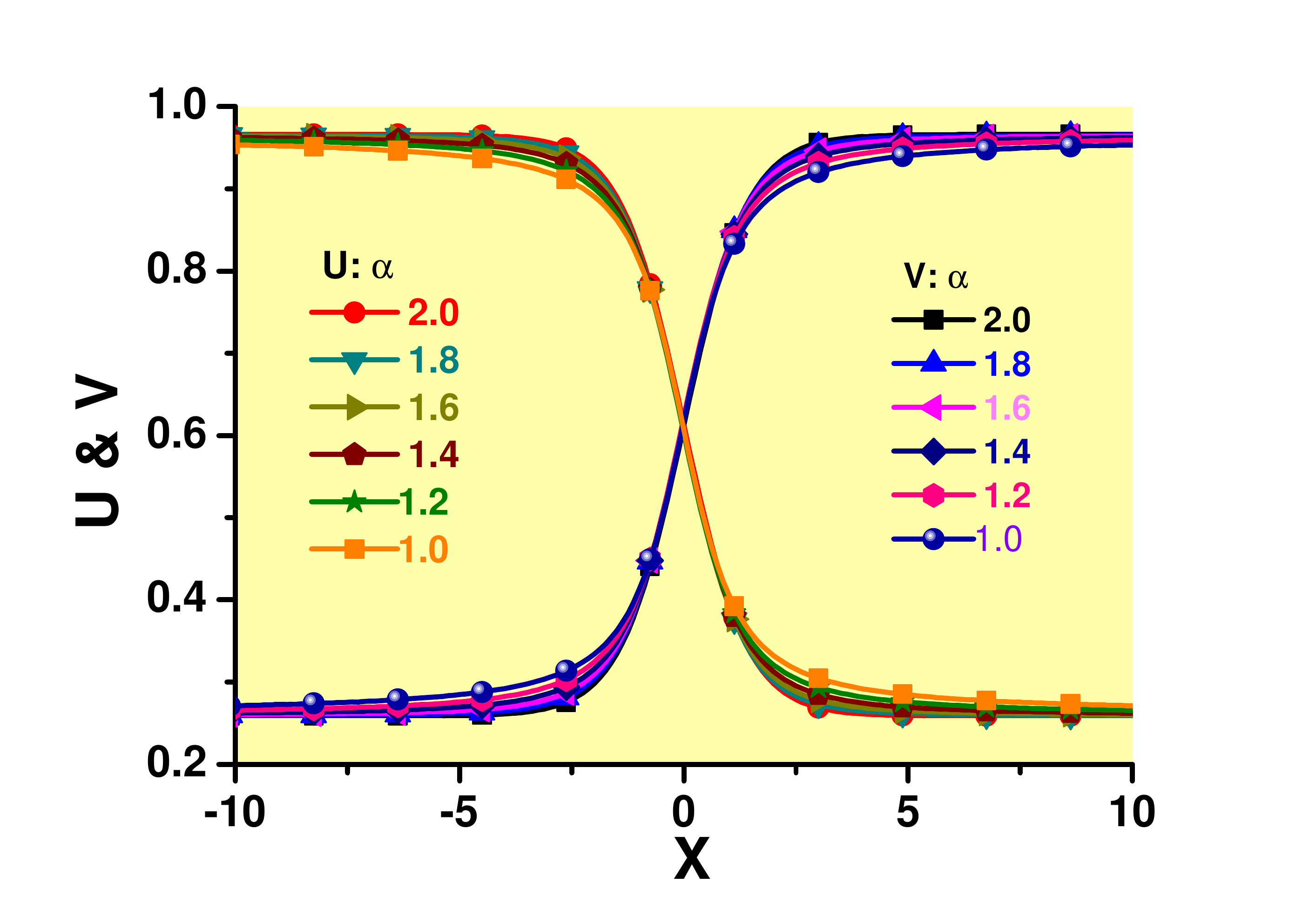}} \subfigure[]{%
\includegraphics[scale=0.3]{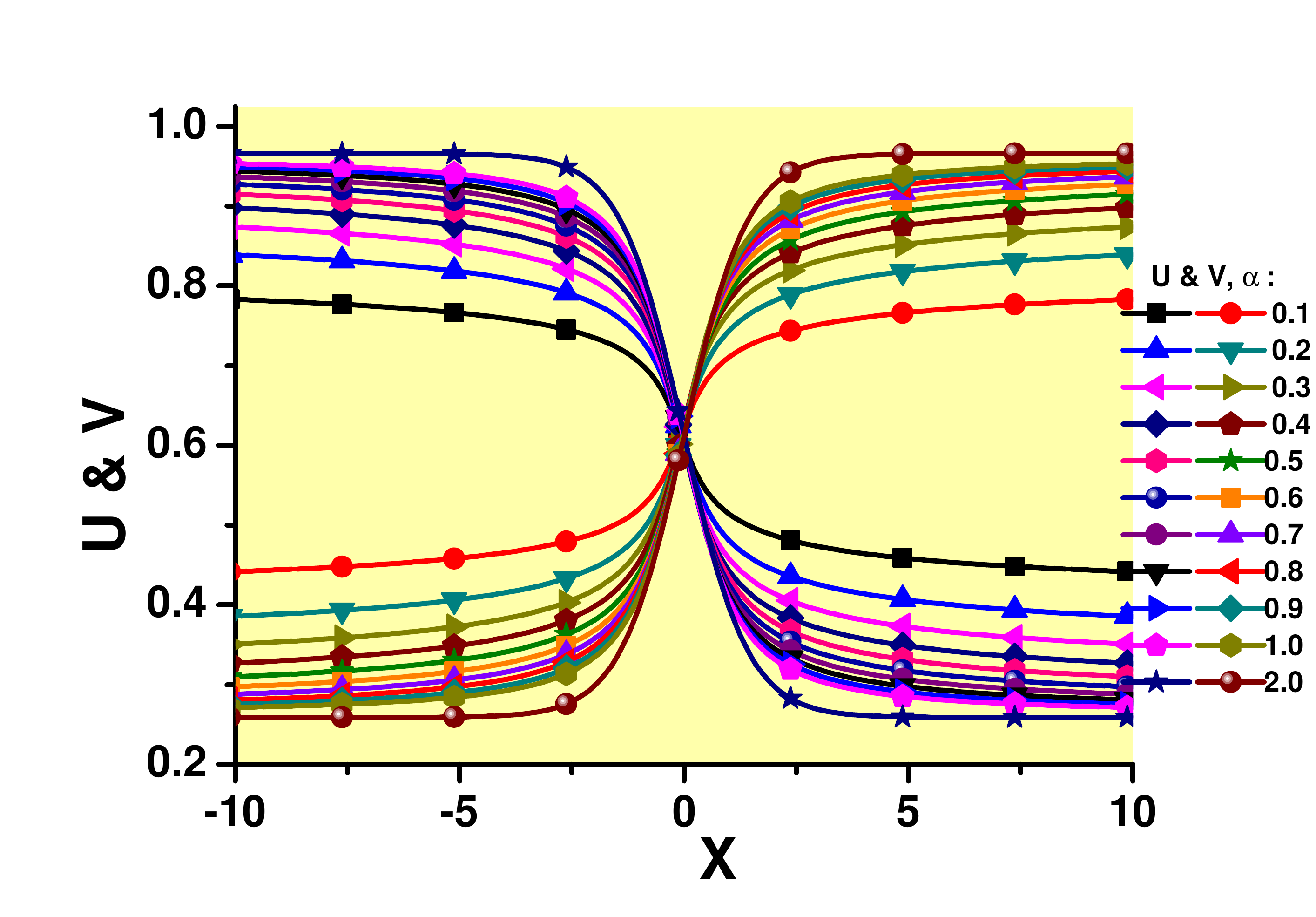}}\newline
\subfigure[]{\includegraphics[scale=0.3]{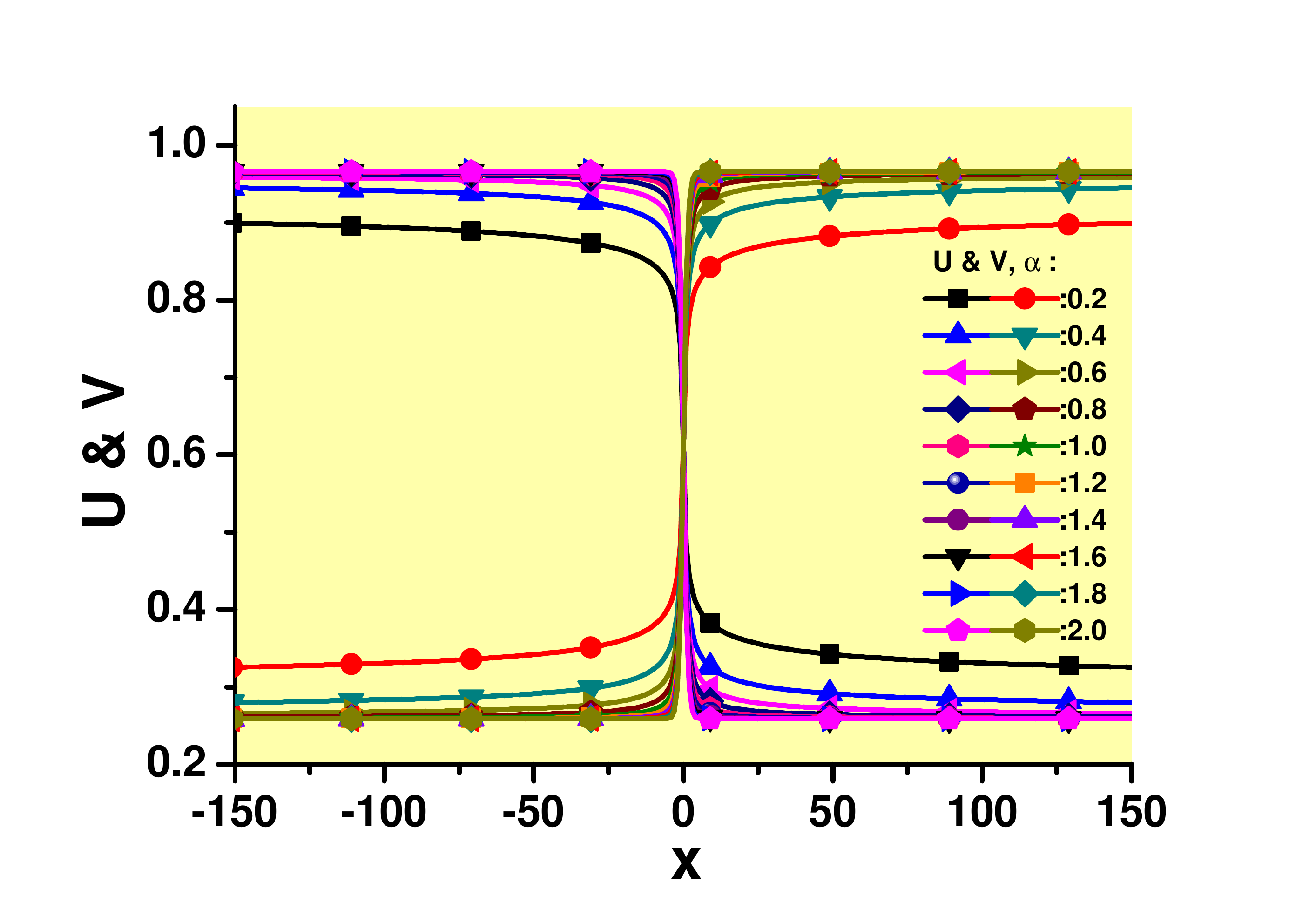}}
\caption{Shapes of stable DW solutions produced by Eqs. (\protect\ref{ansatz}%
) and (\protect\ref{W}) for $\protect\beta =3$, $\protect\lambda =0.5$, $k=1$%
, and the LI varying in intervals indicated in the panels. (a) Narrow DWs
for $1\leq \protect\alpha \leq 2$; (b) broad DWs for $0.1\leq \protect\alpha %
<1$ [for the comparison's sake, the DW profile with $\protect\alpha =2$,
i.e., in the case of the normal diffraction, is also included in (b)]; (c)
both narrow and broad DWs, displayed in a much larger spatial domain for the
entire range of the values of the LI, $0.2\leq \protect\alpha \leq 2$.}
\label{fig5}
\end{figure}

The inclusion of the linear coupling in Eq. (\ref{UV}) neither destroys DW
solutions nor destabilizes them, but makes their shapes more complex, even
in the case of $\beta =3$, when substitution (\ref{ansatz}) reduces the
system of two equations (\ref{UV}) to the single equation (\ref{W}). First,
Fig. \ref{fig5}(a) demonstrates that the linear coupling with strength $%
\lambda =0.5$ produces a relatively weak effect on the shape of the DW
states if the LI takes values in the interval of $1\leq \alpha \leq 2$. On
the other hand, Fig. \ref{fig5}(b) demonstrates a more conspicuous effect of
the same linear coupling for smaller values of the LI, \textit{viz}., in the
interval of $0.1\leq \alpha <1$: the corresponding DWs become essentially
broader, in comparison to their counterparts found at $\lambda =0$. These
results are summarized by Fig. \ref{fig5}(c), where, in a broader spatial
domain, it is shown that the DW solutions eventually converge to asymptotic
values at $|x|\rightarrow \infty $, which are, according to Eqs. (\ref{lim})
and (\ref{ansatz}),
\begin{equation}
U_{\pm }=\left( \sqrt{0.725}\pm \sqrt{0.225}\right) /\sqrt{2}\approx \left\{
0.94,0.27\right\}  \label{U+-}
\end{equation}%
for the case of $k=1$, $\lambda =0.5$, $\beta =3$, although at small values
of $\alpha $ the convergence is very slow. These findings are readily
explained by the scaling relation (\ref{L}), which predicts the growth of
the DW's width with the increase of the linear-coupling constant $\lambda $
and decrease of the LI, $\alpha $.

\begin{figure}[tbp]
\begin{center}
\includegraphics[width=0.60\textwidth]{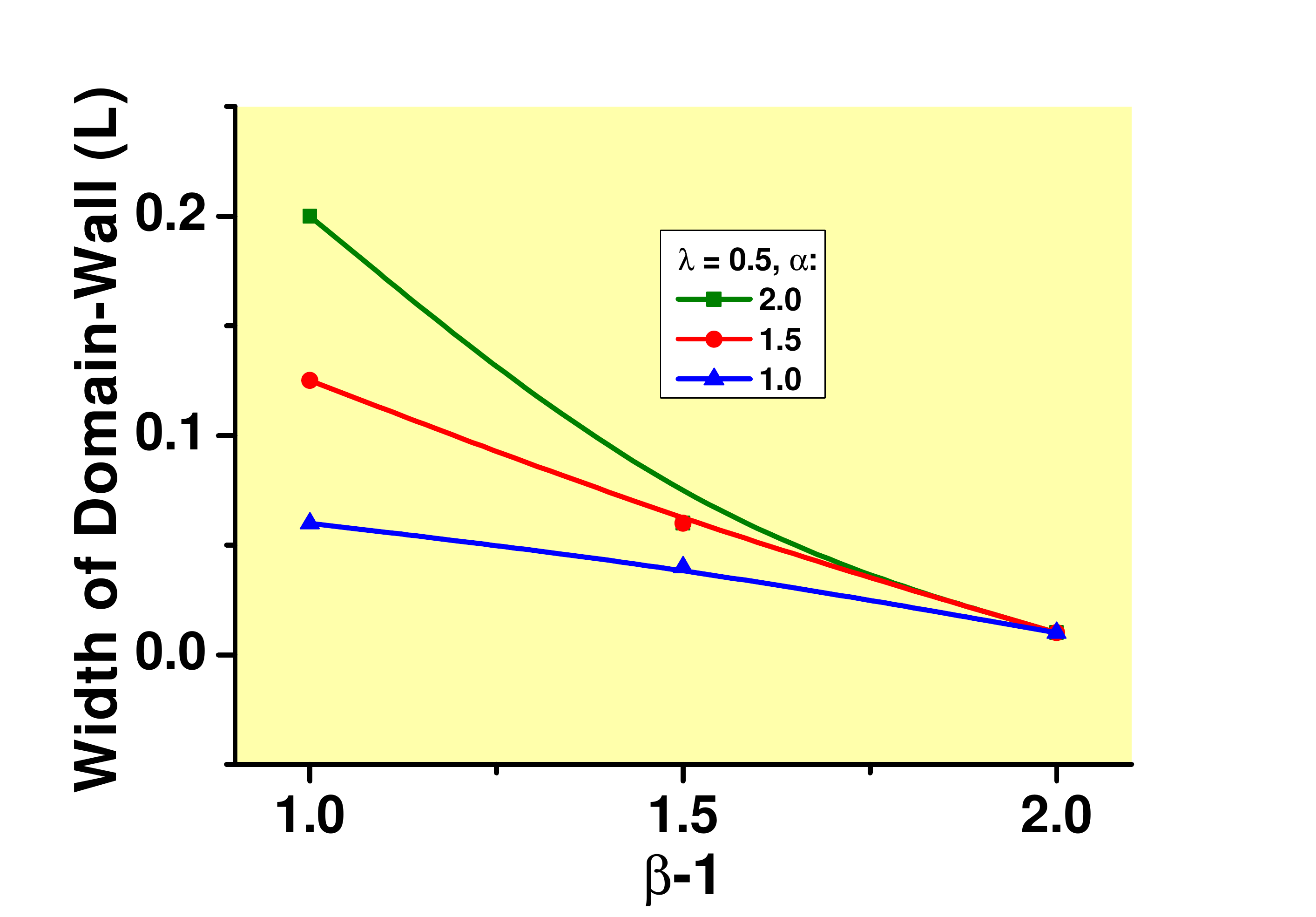}
\end{center}
\caption{The numerically found dependences of width $L$ of the DW on $\left(
\protect\beta -1\right) $ for $k=-1$, $\protect\lambda =0.5$, and three
different values of the LI, $\protect\alpha =2$ (which corresponds to the
normal non-fractional diffraction), $\protect\alpha =1.5$, and $\protect%
\alpha =1$. }
\label{fig6}
\end{figure}

The effects of the linear coupling on the DW states at $\beta \neq 3$ are
qualitatively similar to those displayed in Fig. \ref{fig5} for $\beta =3$.
In all the cases, the DW solutions remain stable if the linear coupling is
incorporated. Further, for comparison of the $L(\beta -1)$ dependences which
are displayed for $\lambda =0$ in Fig. \ref{fig4}, similar dependences for $%
\lambda =0.5$ are presented in Fig. \ref{fig6}. In this case, the
dependencies end close to $\beta -1=1$, in accordance with Eq. (\ref{max}),
which gives $\left( \beta -1\right) _{\mathrm{immisc}}=1$ for $\lambda =0.5$
and $k=-1$. The fact that all the curves yield the same DWs' width at $\beta
=3$ is explained by the above finding that this value plays a special role,
simplifying the DW solutions.

\subsection{DWs in a system with unequal diffraction coefficients and
different values of the L\'{e}vy index}

A relevant generalization of the system of coupled equations which give rise
to DW solutions is one with different diffraction coefficients, $D_{1}\neq
D_{2}$, in the two equations, as recently proposed in Ref. \cite{PLA} (in
the case of the normal diffraction). In the optical system, unequal
coefficients $D_{1,2}\equiv \cos ^{2}\theta _{1,2}$ are determined by
different angles $\theta _{1,2}$ between carrier wave vectors of the two
components of the light waves and the common propagation direction. The
effective diffraction coefficients are also unequal in the BEC model
including two heteronuclear components with different atomic masses, $%
m_{1,2}\sim 1/D_{1,2}$, but in the latter case only the system with $\lambda
=0$ is a physically relevant one (two different atomic species cannot
transform into each other).

\begin{figure}[tbp]
\begin{center}
\includegraphics[width=0.60\textwidth]{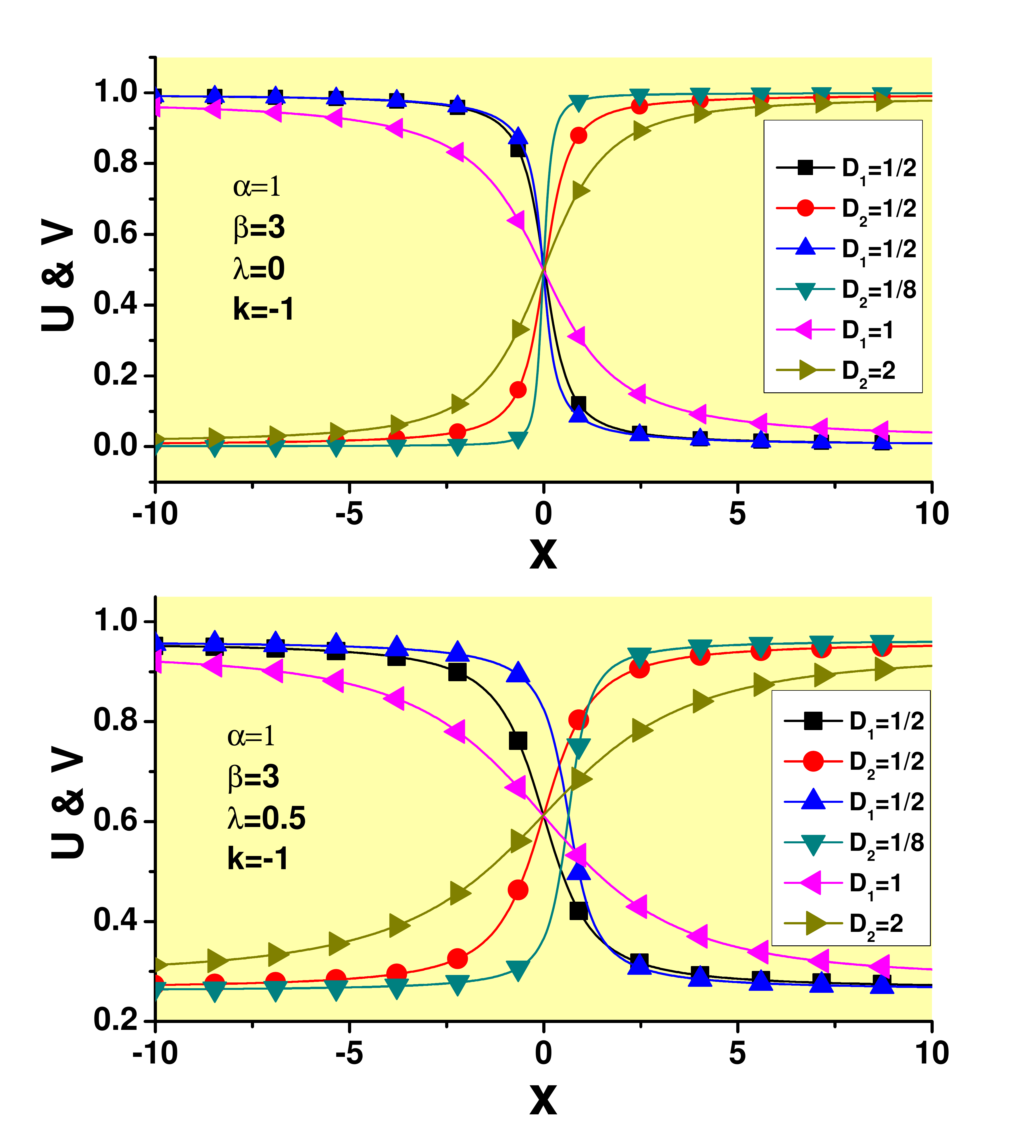}
\end{center}
\caption{Examples of stable asymmetric DWs with $D_{1}\neq D_{2}$, produced
by the numerical solution of Eq. (\protect\ref{DD}) with $\protect\alpha =1$%
, $\protect\beta =3$, $k=-1,$ and two different values of the
linear-coupling constant, $\protect\lambda =0$ and $\protect\lambda =0.5$ in
the top and bottom panels. respectively. The two field components, $U$ and $%
V $, are displayed for three pairs of values $D_{1,2}$, which are indicated,
from top to bottom, in the panels.}
\label{fig7}
\end{figure}

The extension of Eq. (\ref{UV}) with $D_{1}\neq D_{2}$ takes the form of
\begin{eqnarray}
kU+\frac{D_{1}}{2}\left( -\frac{\partial ^{2}}{\partial x^{2}}\right)
^{\alpha /2}U+(U^{2}+\beta V^{2})U-\lambda V &=&0,  \notag \\
kV+\frac{D_{2}}{2}\left( -\frac{\partial ^{2}}{\partial x^{2}}\right)
^{\alpha /2}V+(V^{2}+\beta U^{2})V-\lambda U &=&0.  \label{DD}
\end{eqnarray}%
The numerical solution of Eq. (\ref{DD}) demonstrates that the asymmetry of
the diffraction coefficients makes the shapes of the two components of the
DWs mutually asymmetric, but does not destroy them. An example of the so
deformed shape of the DWs is displayed in Fig. \ref{fig7}, for both cases of
$\lambda =0$ and $\lambda >0$. This solution and all others produced by the
asymmetric system remain completely stable, as shown by the solution of the
respectively modified Eq. (\ref{linearized}), as well as by direct
simulations of the asymmetric version of Eq. (\ref{system}) (not shown here
in detail).

\begin{figure}[tbp]
\begin{center}
\includegraphics[width=0.60\textwidth]{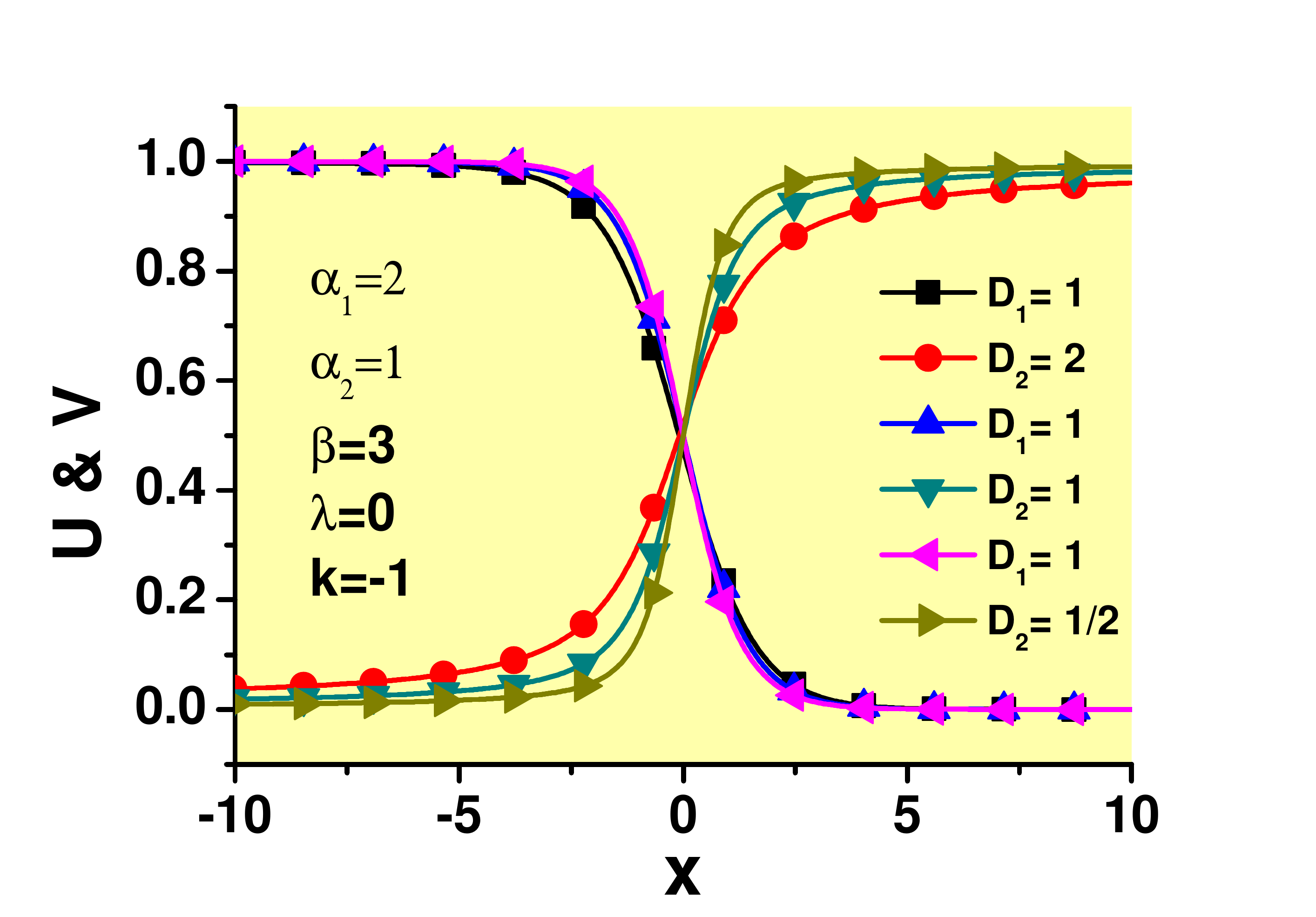}
\end{center}
\caption{Examples of stable asymmetric DWs with different values of the LI
in the two componnts, $\protect\alpha _{1}=2$ and $\protect\alpha _{2}=1$,
produced by the numerical solution of Eq. (\protect\ref{LI}) with $\protect%
\beta =3$, $k=-1,$ and $\protect\lambda =0$. The field components, $U$ and $V
$, are displayed for three pairs of values of the diffraction coefficients, $%
D_{1,2}$, which are indicated, from top to bottom, in the plot.}
\label{fig8}
\end{figure}

Furthermore, it is also possible to consider the system with different
values of the LI, $\alpha _{1}\neq \alpha _{2}$, in the equations for the
two components:
\begin{eqnarray}
kU+\frac{D_{1}}{2}\left( -\frac{\partial ^{2}}{\partial x^{2}}\right)
^{\alpha _{1}/2}U+(U^{2}+\beta V^{2})U-\lambda V &=&0,  \notag \\
kV+\frac{D_{2}}{2}\left( -\frac{\partial ^{2}}{\partial x^{2}}\right)
^{\alpha _{1}/2}V+(V^{2}+\beta U^{2})V-\lambda U &=&0,  \label{LI}
\end{eqnarray}%
where we set $\alpha _{1}=2$ (non-fractional diffraction) and $\alpha _{2}=1$%
. Such a system may be realized in terms of BEC, considering immiscible
components which represent usual particles ($U$) and those moving by the L\'{%
e}vy flights ($V$). This system also supports stable DW states, as shown in
Fig. \ref{fig8}.

\section{Conclusion}

The objective of this work is to demonstrate that the variety of states
maintained by the interplay of the fractional diffraction and
self-defocusing cubic nonlinearity, which may be realized in optics and BEC,
can be expanded by predicting stable DWs (domain walls) in the two-component
system of immiscible fields. The numerical results clearly demonstrate that,
in the entire range of values of the respective LI (L\'{e}vy index), $\alpha
<2$, that determines the fractional diffraction, and at all values of the
relative XPM/SPM coefficient $\beta $, which exceed the immiscibility
threshold, given by Eq. (\ref{max}), the DWs exist and are stable. The same
is true for DWs in the system including the linear mixing between the
components, in addition to the XPM interaction between them (which makes the
immiscibility incomplete). The main characteristic of DW structures is their
width. The present analysis demonstrates that the fractional diffraction
essentially affects the scaling which determines the dependence of the width
on the system's parameters. Numerical results for the scaling corroborate
analytical findings, represented by Eqs. (\ref{L}), (\ref{WDS}), and (\ref%
{beta-1}). It is seen that the decrease of $\alpha $ leads to steep increase
of the scaling exponents $\sim 1/\alpha $, which is explained by the fact
that the fractional diffraction is represented by the nonlocal operator [the
Riesz derivative, defined as per Eq. (\ref{Riesz})]. It is also demonstrated
that the DW solutions are essentially simplified in the special case of $%
\beta =3$. Currently available techniques should make it possible to create
the predicted DWs patterns in the experiment that may be performed for the
bimodal light propagation in the temporal domain \cite{arXiv}.

As extension of the present analysis, it may be natural to study static and
dynamical states built as multi-DW patterns. Another relevant direction is
the consideration of two-dimensional settings, such as radial DWs (also in
static and dynamical states) between immiscible light waves in bulk
waveguides, cf. Refs. \cite{DOPO}-\cite{Bookman}.

\section*{Acknowledgment}

This work was supported, in part, by the Israel Science Foundation through
the grant No. 1695/22.

\end{document}